\documentstyle[prl,aps,preprint,psfig,epsfig]{revtex}
\draft
\tightenlines
\newcommand{\be}{\begin{equation}}
\newcommand{\ee}{\end{equation}}
\newcommand {\vs} {\vskip .5 true cm}
\begin{document}
\title{Electron Decay}
\author{T. Pradhan} 
\address{Institute of Physics, Bhubaneswar 751 005, Orissa, India \\
e-mail: pradhan@iopb.res.in}
\maketitle{}

\begin{abstract} 
The electron would decay into a photon and neutrino if the law of electric
charge conservation is not respected. Such a decay would cause vacancy in closed
 shells of atoms giving rise to emission of x-rays and Auger electrons.
Experimental searches for such very rare decay have given an estimate for the
 life time to be greater than $2.7 \times 10^{23}$ years. The simplest 
theoretical model which would give rise to such a decay is one where
the electron is regarded as the first excited state and neutrino as the
 ground state of a fundamental spin 1/2 particle bound to a scalar particle
 by a super strong force and the photon is considered as a bound state of a
fundamental spin 1/2 fermion-antifermion pair. The fine structure 
constant of the super strong coupling is found to be unity from the
masslessness of the neutrino and the lower bound of the mass of the 
fundamental particles is estimated by using quantum mechanical formula for
photon emission by atoms and found to be $10^{22}$ GeV from the bound for
 electron decay time indicating thereby that the composite nature of electron,
 neutrino and the photon would be revealed in the Planckian energy regime.
A model based on extension of $SU(2)\otimes SU(2)$ symmetry of Dirac
equation to $SU(3)\otimes SU(3)$ gives a lower bound for the mass of the
gauge boson mediating the decay to be $10^9 GeV$  which is the geometric mean
 of the masses of the electron and the fundamental particles. 

\noindent
{PACS -13.30 -r, 13.90 +i, 14.60 -z}
\end{abstract}
\section{Electron Decay}
\vs
The stability of the proton constituting the nucleus of the lightest and most
abundant element, the hydrogen atom, has been the topic of extensive theoretical
 and experimental investigations in the last three decades [1]. However, not
much attention has been paid to the stability of its partner, the electron.
Experiments give a lower bound of $2.7 \times 10^{23}$ years for its life-time
 [2] which is less than that of the proton by several orders of magnitude.

Just as proton would decay, among other particles, to a positron and a neutral
pion if the constraint imposed by baryon number conservation is removed,
so would the electron decay into a photon and a neutrino ( also to a neutrino and
 neutrino-antineutrino pair) if electric charge conservation is not 
respected. Such decay of the electron in closed shells of atoms would cause 
vacancy giving rise to emission of x-rays and Auger electrons. der Mateosian and Goldhaber looked for such
 emission from an Iodine atom located in a NaI(T1) crystal and deduced a 
lower limit of $10^{18}$ years on the electron life time. This limit has been
 revised to $10^{22}$ years for $e\rightarrow \gamma\nu$ by Moe and Reines [3]
 by reducing the detector background. The most recent limit given by the Particle
 Data Group [2] is $2.7\times 10^{23}$ yrs.

The simplest model which would give rise to $e\rightarrow \gamma\nu$ decay is
 one where the electron is regarded as the first excited state and the neutrino
as the ground state of a fundamental spin 1/2 particle bound to a scalar particle
 by a super strong force. Since electric charge is not to be conserved,
 and therefore not a good quantum number, the fundamental particles of the
model should not possess electric charge. Consequently the photon can 
not couple to them directly. It can do so through its constituents if it
 too is regarded as a composite particle. It may be recalled in this connection
 that long ago Fermi and Yang [4] constructed a model for the composite
pion and eta particles according to which
$$\matrix {\phi_i & = N^+\tau_iN\cr
 \eta & = N^+N\cr}\eqno{(1)}$$
where
$$ N = (\matrix{p\cr n\cr})$$
and $\tau_i$ are isotopic spin matrices. The pion field $\phi_i$ is a vector
 in isospin or SU(2) space; the eta $(\eta)$ the nucleon (N) are scalar and
spinors respectively in this space. In terms isospin up (p), isospin down
(n), the pion field as defined, in (1) can be written as
$$\matrix{\phi^{(+)} = &  n^+p\cr
 \phi^{(-)} = &  p^+n\cr
\phi^{(0)} = &  {1\over\sqrt 2}(p^+p- n^+n)\cr
\eta \ \ \ = &  {1\over\sqrt 2}(p^+p+ n^+n)\cr}\eqno{(2)}$$
In the case of photon we have to consider four dimensional space-time instead
 of two-dimensional isospin space and construct composite photon states from
 spin up and down states of fundamental spin 1/2 particles, which span a
$SU(2)\otimes SU(2)$ space. The Dirac equation for these particles can be written in a
 symmetrical from using spinor representation of Dirac matrices :
$$\matrix{(p_0-\vec\sigma\cdot\vec p)\xi_f & = m\eta_f\cr
(p_0+\vec\sigma\cdot\vec p)\eta_f & = m\xi_f\cr}\eqno{(3)}$$
Here $\sigma_i$ are Pauli spin matrices and $\xi_f$ and $\eta_f$ are two-component
 spinors in terms of which the four component Dirac field $\psi_f$ can be written as
$$\psi_f = (\matrix{\xi_f\cr\eta_f})\eqno{(4)}$$
The spinors $\xi_f$ and $\eta_f$ span the $SU_2(\xi)\otimes SU_2(\eta)$ space.
Analogous to the Fermi-Yang model for pion in isospin spin the photon can be
 represented in $SU_2(\xi)\otimes SU_2(\eta)$ space as [5]
$$\matrix{ A_R = &  \psi_f^{(d)^+}\psi_f^{(u)}\cr
 A_L =  &  \psi_f^{(u)^+}\psi_f^{(d)}\cr
 A_3 = & {1\over\sqrt 2} ( \psi_f^{(u)^+}\psi_f^{(u)}-\psi_f^{(d)^+}\psi_f^{(d)})\cr
 A_0 = & {1\over\sqrt 2} ( \psi_f^{(u)^+}\psi_f^{(u)}+\psi_f^{(d)^+}\psi_f^{(d)})\cr}\eqno{(6)}$$
where u and d stand for spin s up and down, $A_R$ and $A_L$ represent right and left
 circularly polarized photon, and $A_3$ and $A_0$ represent longitudinal and
 time-like photon fields. It will be noticed that the $A_R$ and $A_L$ are 
composed of fundamental fermions with parallel spins, $A_3$ and $A_0$ are composed
of those with antiparallel spins. Just as parallel electric currents attract and
 antiparallel ones repel, so would parallel spins attract and antiparallel
 spins repel [6] since the force is of gauge origin as in quantum electrodynamics.
 For this reason $A_R$ and $A_L$ will be bound states while $A_3$  and $A_0$ remain
 unbound. This explains why there only two physical photon states.

Since the electron and neutrino have spin 1/2 and our fundamental fermions have
also have spin 1/2, they former should be of composites of such a fermion and a 
fundamental spinless particle which is the counter part of the strange quark in
 SU(3) space, electron and neutrino being counterparts of K-mesons. In order to
 accommodate the spinless field we have to extend the $SU_2\otimes SU_2$ model
 to $SU_3\otimes SU_3$ and use extended Dirac equation :
$$\matrix{(p_0-\lambda_ap_a)\xi  & = m\eta\cr
(p_0+\lambda_ap_a)\eta  & = m\xi\cr}\eqno{(7)}$$
where
$$\xi = \big (\matrix{\xi_1\cr \xi_2\cr \xi_s\cr}\big ) \ \ 
\eta = \big (\matrix{\eta_1\cr \eta_2\cr \eta_s\cr}\big )\eqno{(8)}$$
and $\lambda_a$(a=1,2,....8) are $(3\times 3) \ SU_3$ matrices. These two 
equations can be combined into a single one :
$$(\Gamma_0 p_0-\Gamma_ap_a) \psi = m \psi\eqno{(9)}$$
where
$$\psi = \big (\matrix{\xi\cr\eta\cr}\big ) \ \ \Gamma_0 = \big (\matrix{0 & 1\cr
1& 0\cr}\big ) \ \ \Gamma_a = \big (\matrix{0 & -\lambda_a\cr \lambda_a &
0\cr}\big )\eqno{(10)}$$
which can be derived from the Lagrangian density
$$ {\cal L} = \overline\psi(\Gamma_0p_0-\Gamma_ap_a)\psi-m\overline\psi\psi\eqno{(11)}$$
The gauge interaction incorporated by replacing $p_0$ by $(p_0-gG_0)$ and
$(p_a-gG_a)$ in (11) gives
$${\cal L}_{int} = - g\overline\psi\Gamma_0G_0\psi-g\overline\psi\Gamma_a p_a
\psi\eqno{(12)}$$
The interaction between the  fundamental spin 1/2 and spin 0 fields 
relevant for the formation and decay
 of the electron and neutrino contained in this comes out to be
$${\cal L}_{fs}=-g\psi_f^+\gamma_5\psi_sG_k\eqno{(13)}$$
where $ \gamma_5 = \big (\matrix{-1 & 0\cr 0 & +1\cr}\big )$ \ \ and \ \ 
$G_k=\big(\matrix{G_4 -i& G_5\cr G_6-i & G_7\cr}\big)$
$$ \psi_f = \big (\matrix{\xi_f\cr \eta_f\cr }\big ) \ \ 
\psi_s = \big (\matrix{\xi_s\cr \eta_s\cr}\big )$$
The electron and neutrino are first excited and ground states of the 
composite
$$\matrix{e^{(u)} = \psi^+_s\psi_f^{(u)} & e^{(d)} = \psi_s^+\psi_f^{(d)}\cr
\overline e^{(u)} = \psi^{(u)}_f\psi_s^+ & \overline e^{(d)} = \psi_f^{(d)}\psi_s^+\cr}
\eqno{(15)}$$

The interaction (13) leading to electron decay can be represented by the
Feynman diagram 
\begin{figure}[hbt]
\begin{center}
\includegraphics[width=3in]{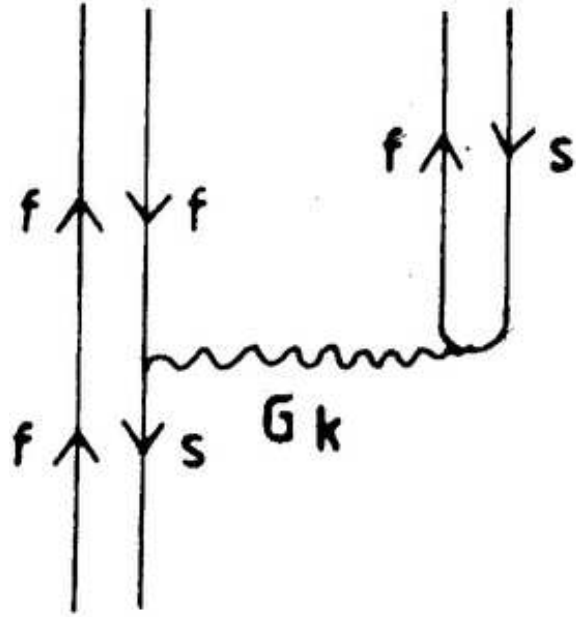}
\end{center}
\caption{Feynman diagram for Electron decay}
\end{figure}
\noindent from which the life-time of the decay (as in proton decay) comes out to be
$${1\over\tau} = \alpha^2_g {m_e^5\over m_k^4}\eqno{(16)}$$
Taking $\alpha_g=1$ and using the experimental lower bound for $\tau$ as
$$\tau > 2.7 \times 10^{21} \ years \eqno{(17)}$$
gives the bound for $m_k$ as
$$m_k >  10^9 \ Gev \eqno{(18)}$$

A lower bound for the mass of the fundamental particles can be obtained by making
 use of the formula
$${1\over\tau} = {4w^3\over 3}\mid < \nu\gamma \mid M\mid e>\mid^2\eqno{(19)}$$
used for atomic transitions since the electron and neutrino have been taken to
be excited and ground state of spin 1/2 composites of fundamental spin 1/2and 
spin zero particles. In this formula, in that case, $\omega$ would equal the electron
mass. For obtaining an estimate, we take
$$ <\nu\gamma\mid M\mid e> = \alpha_g/m\eqno{(20)}$$
Taking $\alpha_g = 1$ we get from (19) and (20)
$$\tau m^2/m^3_e\eqno{(21)}$$
This gives the bound on m as
$$m > 10^{22} Gev\eqno{(22)}$$
Combining (21) and (16) gives 
$$m_em=m^2_k\eqno{(23)}$$

Having considered electron decay we take up the problem of formation of electron through exchange
 of massive gauge bosons $G_k$ between $\psi_s \& \ \psi_f$. The second 
order Dirac equation for this can be written
$$[{d^2\over dr^2}+{2\over r} - {\Lambda^2\over r^2} + 2\alpha_g E{e^{im_kr}\over r} -
({m^2\over 4}-E^2)]R(r)=0\eqno{(24)}$$
where
$$\Lambda^2 = \lambda(\lambda+1) , \ \mid \lambda \mid = \sqrt{(j-{1\over 2})^2-
\alpha_g^4}$$
In terms of scaled variables
$$\rho = r/\beta \ \ \in = E\beta \ \ \kappa = {\beta m\over 4}, \ \ (\beta <<1)\eqno{(25)}$$
this equation reads
$$[{\partial^2\over\partial\rho^2} +{2\over \rho}{d\over d\rho}+{\Lambda^2\over\rho^2}
- {2\alpha_g\in e^{-m_k\beta\rho}\over \rho}+(4K^2-\in^2)]R=0\eqno{(26)}$$
Since $\beta < < 1$ we can put $e^{-m_k\beta\rho}=1$ in which case equation (24)
becomes a second order Dirac equation with Coulomb potential whose solution for
 energy $\in$ is
$$\in = {2m_e\over\sqrt{1+{\alpha_g^2\over (\sqrt{\kappa^2-\alpha^2}+n_r)^2}}}\eqno{(27)}$$
where
$$\kappa = \cases{l \ for \ j = l+{1\over 2} \hfil&\hfil\cr
-(l+1) \ for \ j = l-{1\over 2}}\ \  and \ \ n_r = n-\mid\kappa\mid.$$
 For the ground
state which we have taken as the neutrino, we get from this
$$m_{\nu} = 2m_2\sqrt{1-\alpha^2_g}\eqno{(28)}$$
i.e., 
$$\alpha_g = \sqrt{1-({m_{\nu}\over 2m_2})^2}$$
Similarly for the electron, which we have taken to be the first excited state,
 we get 
$$m_e = \kappa +{m_{\nu}\over 2}\eqno{(29)}$$
If we take $m_{\nu}=0. \ \alpha_g=1, \ \kappa = m_e$ in wich case $\beta \sim
10^{-25}$ as assumed. It will be noted that while the mass and energy in
 eqn(23) are in the Planck regime, those in eqn(25) are in the atomic regime.

We thus see that a model where the electron and neutrino aare taken to be
excited and ground states of a composite of a fundamental spin 1/2 fermion
 and a spin zero particle and the photon as a composite of the fundamental
 fermion antifermion pair, leads to relation between electron decay time, mass
 of the fundamental particles and also the mass of the gauge boson which provides
 the binding. The experimental bound for electron life time gives bounds for these
 masses. While the mass of the fundamental particles lies in the Planckian regime,
 the mass of the gauge boson is the geometric mean of this mass and that of the 
electron.

\end{document}